\documentclass[twocolumn,showpacs]{revtex4}
\usepackage{graphicx}%
\usepackage{dcolumn}
\usepackage{amsmath}
\usepackage{bm}

\makeatletter
\def\btt#1{\texttt{\@backslashchar#1}}%
\DeclareRobustCommand\bblash{\btt{\@backslashchar}}%
\makeatother

\begin{document}

\preprint{HEP/123-qed}

\title[Short Title]{Field-Induced Magnetostructural Transitions in Antiferromagnetic Fe$_{1+y}$Te$_{1-x}$S$_x$}%

\author{M.~Tokunaga$^{1,2}$}
 \email{tokunaga@issp.t.u-tokyo.ac.jp}
\author{T.~Kihara$^1$}%
\author{Y.~Mizuguchi$^{2,3}$}%
\author{Y.~Takano$^{2,4}$}%

\affiliation{%
$^1$ The Institute for Solid State Physics (ISSP), The University of Tokyo, Kashiwa 277-8581, Japan \\
$^2$ JST-TRiP, Chiyoda, Tokyo 102-0075, Japan\\
$^3$ Tokyo Metropolitan University, Hachioji, Tokyo 192-0397, Japan\\
$^4$ The National Institute for Materials Science, Tsukuba 305-0047, Japan \\
}%

\date{\today}

\begin{abstract}
The transport and structural properties of Fe$_{1+y}$Te$_{1-x}$S$_x$ ($x=0$, 0.05, and 0.10) crystals were studied in pulsed magnetic fields up to 65~T. The application of high magnetic fields results in positive magnetoresistance effect with prominent hystereses in the antiferromagnetic state. Polarizing microscope images obtained at high magnetic fields showed simultaneous occurrence of structural transitions. These results indicate that magnetoelastic coupling is the origin of the bicollinear magnetic order in iron chalcogenides.
\end{abstract}

\maketitle

In 2008, Kamihara $et\ al.$ observed superconductivity below 26~K by partially substituting F for O in the antiferromagnetic metal LaFeAsO~\cite{Kamihara2008}. The occurrence of moderately high temperature superconductivity in the vicinity of magnetic order implies a possible magnetic origin for pairing and hence attracts considerable attention, similar to the case of cuprate superconductors. Subsequent to the study of Kamihara $et\ al.$, a large number of studies have shown the existence of many related superconductors with different crystal structures~\cite{Rotter2008,Wang2008,Ogino2009,Zhu2009,Hsu2008}, and opened a new era in the quest for novel high-temperature superconductors.

As a common structural feature of this class of superconductors, Fe ions located at the centers of tetrahedra of arsenic or chalcogen ions form a layered square lattice network. The electronic states near the Fermi level are characterized mainly by the $3d$ bands of these Fe ions. Since early band calculation for LaFeAsO~\cite{Singh2008} has suggested the presence of cylindrical Fermi surfaces at the ${\rm \Gamma}$ and the M points, the Fermi surface nesting with the wave vector $({\rm \pi},{\rm \pi})$ in the folded Brillouin zone (two Fe ions in the unit cell) has been regarded as the origin of the itinerant antiferromagnetism in these materials. The shape of the Fermi surface experimentally observed by angle-resolved photoemission spectroscopy (ARPES)~\cite{Liu2008} and the antiferromagnetic structure determined by neutron diffraction [Fig.~1(a)]~\cite{delaCruz2008} are consistent with this scenario.

\begin{figure}[!tb]
\includegraphics[width=8.7cm]{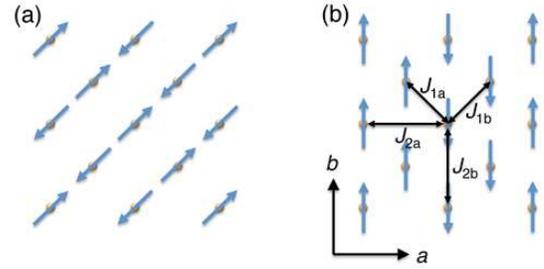}
\caption{(color online) Schematics of the in-plane spin order in the antiferromagnetic states of (a) iron arsenides and (b) iron tellurides. The arrows represent the spins of Fe ions. Exchange interactions of $J_{1a,b}$ and $J_{2a,b}$ were defined as shown in Fig. 1(b).}
\label{fig1}
\end{figure}

On the other hand, Fe chalcogenides pose an additional problem pertaining to our understanding of the magnetism in the parent compounds of Fe-based superconductors. The Fe chalcogenides have been extensively studied because of its simple crystal structure and the availability of large single crystals. The parent material Fe$_{1+y}$Te is an itinerant antiferromagnet with a N$\acute {\rm e}$el temperature ($T_{\rm N}$) of about 70~K. Neutron experiments performed with this material have shown the presence of the bicollinear ($\pi$,0) antiferromagnetic order [illustrated in Fig.~1(b)]~\cite{Li2009}. Since this spin structure is not consistent with the scenario involving nesting between the ${\rm \Gamma}$ and the M points, we have to consider an additional factor to explain this magnetic order. Han $et\ al.$ proposed that the nesting vector can change to ($\pi$,0) when the chemical potential is shifted by the excess Fe ions, which is inevitable in Fe$_{1+y}$Te~\cite{Han2009}. The Fermi surface determined by an ARPES experiment~\cite{Xia2009}, however, is rather similar to those in iron arsenides than that predicted in the theory of Han $et\ al$. Therefore, it is important to clarify the origin of the magnetic order, which may play a crucial role in the occurrence of high-temperature superconductivity in Fe-based superconductors.

In this context, we studied the effect of magnetic fields on the antiferromagnetic state of iron chalcogenides. Application of high magnetic fields is widely recognized as a powerful method to investigate the basic properties of magnetic materials. We studied the transport and structural properties of Fe$_{1+y}$Te$_{1-x}$S$_x$ in high magnetic fields and found the occurrence of field-induced transitions in the antiferromagnetic states.

Single crystals of Fe$_{1+y}$Te$_{1-x}$S$_x$ were grown by the flux method~\cite{Mizuguchi2009,Mizuguchi2011}. The nominal values of $x$ were 0, 0.05, and 0.10, and these corresponded to actual $x_{\rm E}$ values of 0, 0.03, and 0.048 in the electron-probe micro-analyses performed on other sample pieces synthesized in the same conditions~\cite{Mizuguchi2011}. Although we did not evaluate the amount of excess Fe ions, $y$, we chose growth conditions that helped to obtain samples with small $y$ that show the commensurate antiferromagnetic and monoclinic crystal structures below $T_{\rm N}$~\cite{Bao2009}. On these crystals, we measured the dc and ac magnetoresistance in high magnetic fields up to 65~T using non-destructive pulse magnets installed at The Institute for Solid State Physics. Structural changes in high magnetic fields were studied by using the newly developed high-speed polarizing microscopy system and a small pulse magnet~\cite{Katakura2010}.

\begin{figure}[!tb]
\includegraphics[width=8.7cm]{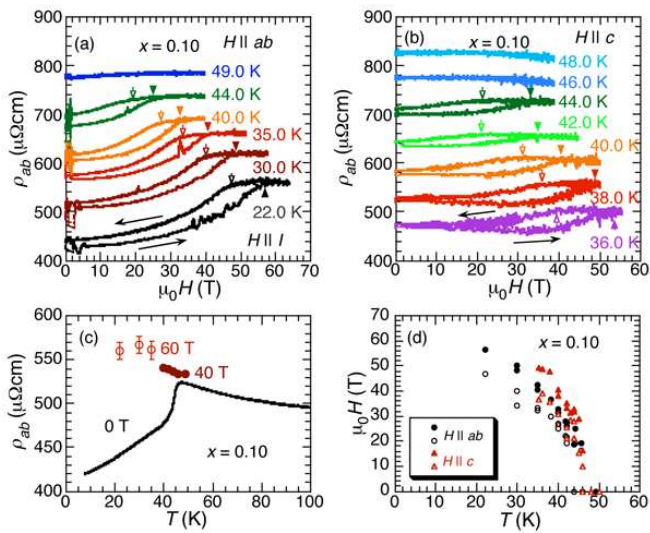}
\caption{(color online) ac magnetoresistance of the $x=0.10$ crystal for (a) $H || ab$ and (b) $H || c$ at various temperatures. The data at $T\geq 30.0$~K were vertically offset by $50\ {\rm \mu \Omega}$cm for clarity. (c) Temperature dependence of the in-plane resistivity at $\mu _0H=0$, 40, and 60~T. (d) The phase diagram for $x=0.10$ in the $H$-$T$ plane. Solid (open) circles and triangles represent transition fields in the field increasing (decreasing) process for $H || ab$ and $H || c$, respectively.}
\label{fig2}
\end{figure}

Figure 2(a) shows the longitudinal in-plane magnetoresistance of the $x=0.10$ crystal at various temperatures ($T$), measured at a frequency of 50~kHz. This figure was plotted with vertical offsets for clarity. At zero field, the in-plane resistivity ($\rho _{ab}$) shows a steep decrease upon cooling below 46~K, indicating the antiferromagnetic transition at this temperature [Fig.~2(c)]. Below $T_{\rm N}$, the application of a magnetic field ($H$) leads to a gradual increase in $\rho _{ab}$; the increase in $\rho _{ab}$ continues up to a certain field marked by the solid triangle, and $\rho _{ab}$ then saturates. The parameter $\rho _{ab}$ starts to decrease at the field marked by the open triangle in the field decreasing process and shows finite hysteresis. The transition fields marked by these triangles monotonically increase as the temperature decreases from $T_{\rm N}$. Similar magnetoresistance behavior is observed for $H || c$, as shown in Fig.~2(b). We can identify the transition fields, marked by the triangles, although their identification becomes somewhat ambiguous because of the smaller signal-to-noise ratio in this transverse configuration.

Figure~2(d) shows the phase diagram for the $x=0.10$ crystal, obtained by using the magnetoresistance for $H || ab$ (circles) and $H || c$ (triangles). The solid and open symbols represent the transition fields in the field increasing and field decreasing processes, respectively. For a given $T$, the transition field for $H || ab$ is smaller than that for $H || c$. Although we could not reach the transition field at low temperatures, the extrapolation of the phase diagram indicates the transition field to be about 70~T for $H || ab$ in the low temperature limit. As can be observed in Fig.~2(c), the values of $\rho _{ab}$ in the high-field phase lie along the semiconducting $\rho _{ab}$-$T$ curve above $T_{\rm N}$, suggesting the disappearance of the antiferromagnetic order in the high-field phase.

\begin{figure}[!tb]
\includegraphics[width=8.7cm]{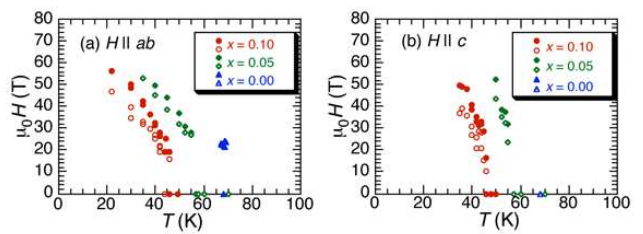}
\caption{(color online) $H$-$T$ phase diagram for (a) $H || ab$ and (b) $H || c$ for $x=0$, 0.05, and 0.10, determined from the experiments on isothermal magnetoresistance.}
\label{fig3}
\end{figure}

Similar first order transitions were observed in the $x=0$ and 0.05 crystals in magnetoresistance measurements. Figs.~3(a) and (b) show phase diagrams for $H || ab$ and $H || c$ for $x=0$, 0.05, and 0.10. Both $T_{\rm N}$ and transition field decrease as $x$ increases. A common feature in the three cases ($x=0$, 0.05, and 0.10) is that the transition fields for $H || c$ are higher that those for $H || ab$.

Now, let us discuss this transition on the basis of the localized spin model. Here, we consider the following Hamiltonian:
\begin{eqnarray}
\mathcal{H} &=& J_z \sum_{i,n}{\bm S}^n_i \cdot {\bm S}^{n+1}_i  + \sum_n \sum_{\langle i,j \rangle}J_{ij}{\bm S}^n_i \cdot {\bm S}^n_j ,
\label{eq:Hamiltonian}
\end{eqnarray}
where $n$ is the layer index. The $J_{ij}$'s denote the in-plane exchange interactions shown in Fig.~1(b). By using a similar local spin model, Fang $et\ al.$ explained the bicollinear spin order shown in Fig.~1(b) by assuming the anisotropic interaction to satisfy $J_{2a} \geq J_{2b}$ and $J_{1a} \geq J_{1b}$~\cite{Fang2009}. In this model, the energy of the bicollinear antiferromagnetic state per Fe ion is given by
\begin{eqnarray}
E_{\rm b-AFM} &=& - J_{1a} - J_{2a} + J_{1b} + J_{2b} - J_z.
\end{eqnarray}
Here,  we assume $S=1$ for the Fe spins. On the other hand, the energy in the ferromagnetic state is given by
\begin{eqnarray}
E_{\rm FM} &=& J_{1a} + J_{2a} + J_{1b} + J_{2b} + J_z.
\end{eqnarray}
From inelastic neutron experiments on Fe$_{1.05}$Te~\cite{Lipscombe2011}, the exchange constants were evaluated as $J_{1a}=-17.5$~meV, $J_{1b}=-51.0$~meV, $J_2=J_{2a}=J_{2b}=21.7$~meV, and $J_z=1$~meV~\cite{J3}. Using these values, the energy difference between the two states was estimated to be 10.4~meV. This energy difference is overcome by the Zeeman energy at a field of 90~T for $g=2$ and $S=1$, which roughly coincides with the transition fields we observed. If we consider both $J_{1a}$ and $J_{2a}$ to be positive, as expected for realizing the bicollinear state in the local spin model, then the transition field cannot be as small as the field we observed. In this respect, our present results are consistent with the neutron study that showed the existence of competition between $J_{1a}$ and $J_{2a}$.

The experimentally determined set of $J_{ij}$'s had an isotropic $J_2$, which cannot reproduce the bicollinear spin order. According to a theoretical study~\cite{Turner2009}, anisotropy in the exchange constant is associated with the orbital order. When the degeneracy of the $d_{zx}$ and $d_{yz}$ orbitals is lifted by the Jahn-Teller effect, antiferromagnetic exchange interaction in the orbital direction promotes the formation of antiferromagnetic chains. Interchain coupling is thought to be caused by the ferromagnetic double exchange interaction mediated by carriers introduced by the excess Fe ions. In Ref.~20, the authors stated that if the orbital order determines the spin order, structural distortion can persist even if the antiferromagnetic order is removed by external magnetic fields. To study the structural properties, we performed high-speed polarizing microscopy in pulsed high magnetic fields.

\begin{figure}[!tb]
\includegraphics[width=8.7cm]{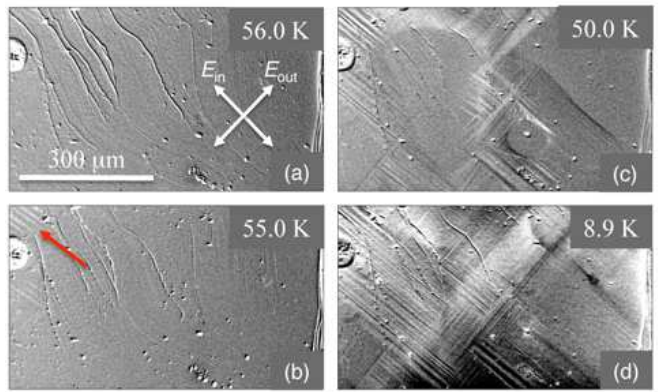}
\caption{(color online) (a)-(d) Differential polarizing microscope images of the cleaved $ab$-plane surface of the $x=0.05$ crystal at various temperatures in a cooling process. The arrows in (a) represent the polarization directions of the incident ($E_{\rm in}$) and reflected ($E_{\rm out}$) lights.}
\label{fig4}
\end{figure}

In Fig.~4, we show polarizing microscope images of the cleaved $ab$-plane of the $x=0.05$ crystal at various temperatures in zero field. These images were captured by a cooled charge-coupled-device camera. This sample shows a transition from a tetragonal to a monoclinic structure upon cooling at $T_{\rm N}=56$~K. This structural change breaks the fourfold symmetry in the $ab$-plane and results in twin domains that show up as stripe-like structures in polarizing microscope images~\cite{Tanatar2009}. To emphasize the change caused by the structural transition, the background image taken at 57~K was subtracted from the images in Fig.~4. Upon cooling, stripe-like structures appeared at 55~K around the left-top corner of the visual area in Fig.~4(b) (marked by an arrow). With a further decrease in the temperature, this stripe-like feature gradually spreads throughout the whole area of the sample [Figs.~4(c) and (d)]. The series of images can be seen in the supplemental movie. This gradual change is not likely to originate from chemical inhomogeneity since our SEM-EDX analysis of this sample piece does not show such features.

\begin{figure}[!tb]
\includegraphics[width=8.7cm]{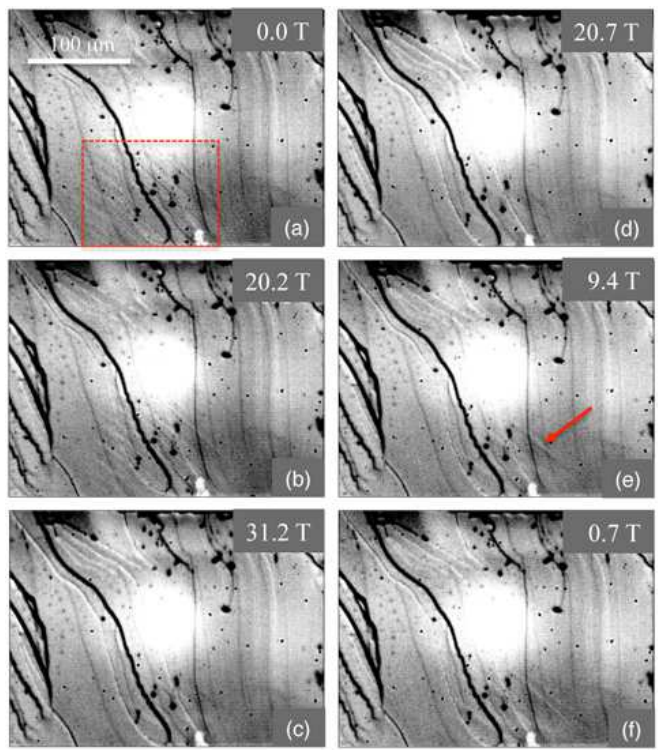}
\caption{(color online) Polarizing microscope images of the $x=0.05$ crystal at $T = 50$~K captured by using a high-speed camera for pulsed magnetic fields up to about 31~T applied normal to the surface. The series of images are shown in the supplemental movie.}
\label{fig5}
\end{figure}

If the hysteretic change in magnetoresistance is coupled with a structural transition, we will see the disappearance of the stripe-like structures observed above in magnetic fields. We obtained polarizing microscope images in pulsed fields up to 31~T with a high-speed camera at a rate of 6,000 frames per second. The resultant images at 50~K are shown in Figs.~5(a) to (f). The entire set of images can be seen in the supplemental movie. 

In this experiment, the relative angle between the polarization directions of the incident and reflected lights was set to be far from 90$^\circ$ so as to achieve sufficient brightness even for limited exposure time. Owing to this limitation, the contrast of the stripe-like structure became weaker than those resolved in Fig.~4. However, we can identify the twin domains at zero field in the area marked by the dotted rectangle in Fig.~5(a). The stripe-like structure disappeared upon the application of a magnetic field along the $c$-axis, i.e., normal to the surface, as shown in Figs.~5(b) and (c). In the field decreasing process, this structure partially recovered, not at 20~T (d) but at 9.4~T (e) (marked by an arrow). Such irreversible changes, also seen in (a) and (f), are characteristic of the first-order transition.

This transition cannot be ascribed to domain rotation as reported for Ba(Fe$_{1-x}$Co$_x$)$_2$As$_2$~\cite{Chu2010}, since we observed the changes in $H || c$. Despite the annihilation of the twin domain sets below the transition field observed in Fig.~3(b), it is not surprising that the local structural change occurs at a field lower than that at where the change in resistivity saturates. In addition, some readers may be concerned about the Joule heating caused by the eddy currents induced in pulsed fields. The Joule heating turns out to be negligible when determined through a simple order estimation~\cite{Joule}. Therefore, we conclude that the field-induced hysteretic change in magnetoresistance is accompanied by a structural transition.

This result indicates that the monoclinic distortion, or the orbital order, disappears when the antiferromagnetic order is removed by an applied magnetic field in Fe$_{1+y}$Te$_{1-x}$S$_x$. In addition, the higher resistance in the high-field phase [Fig.~2(c)] seems incompatible with the double exchange mechanism proposed in the theory. Therefore, we should introduce a mechanism other than the orbital order scenario to understand the origin of the unique magnetic order in this system.

Paul $et\ al.$ pointed out the importance of magnetoelastic coupling in the realization of the bicollinear antiferromagnetic order~\cite{Paul2011a,Paul2011b}. According to their theory, the spin order eventually changes from the ($\pi$,$\pi$)-type to the ($\pi$,0)-type as the magnetoelastic coupling increases. Our present results show the presence of significant coupling between the spin and lattice systems in Fe$_{1+y}$Te$_{1-x}$S$_x$. Our previous study showed the similar magnetotransport properties in EuFe$_2$As$_2$~\cite{Tokunaga2010}. Further systematic studies on the spin-lattice coupling in various classes of materials are essential to understand the origin of the magnetism behind the high-temperature superconductivity in Fe pnictides. 

In conclusion, we studied the transport and structural properties of Fe$_{1+y}$Te$_{1-x}$S$_x$ in high magnetic fields. The results indicate the occurrence of field-induced transitions that can be ascribed to the collapse of the antiferromagnetic order and concomitant structural transitions to the tetragonal phase.  These findings point to the importance of magnetoelastic coupling, which can be a key factor in the realization of the unique magnetic order in Fe chalcogenides.

\begin{acknowledgments}
This work was supported by the MEXT, Japan, through Grant-in-Aid for Scientific Research (23340096).
\end{acknowledgments}


\begin{thebibliography}{}
\bibitem{J3} We neglected the third neighbor interaction $J_3$, bacause Lipscombe $et\ al.$ reported that the low energy magnetic dispersion can be reproduced without using the $J_3$ [17].
\bibitem{Joule} The total amount of the Joule energy estimated in the sample piece for imaging is 300~nJ, which will cause the heating of 5~mK in the adiabatic condition. In the real experiments, heat transfer to the thermal bath significantly diminishes the heating.
\end{thebibliography}
\end{document}